# EVALUATING PATIENT SAFETY RISKS IN GENERATIVE AI: DEVELOPMENT AND VALIDATION OF A FMECA FRAMEWORK FOR GENERATED CLINICAL CONTENT

Lydie Bednarczyk MD [1,2,‡], Jamil Zaghir PhD [1,2,*,‡], Julien Ehrsam MD [1,2], Maria Tcherepanova MD [1,2], Christian Skalafouris PhD PharmD [3], Karim Gariani MD PhD PD [4], Catherine Geslin [5], Claire-Bénédicte Rivara MD [6], Pascal Bonnabry PhD [3], Laetitia Gosetto PhD [1,2], Richard Dubos [1,2], Mina Bjelogrlic PhD [1,2], Christophe Gaudet-Blavignac PhD [1,2], Christian Lovis MD MPH FACMI [1,2]

[1] Division of Medical Information Sciences, Diagnostic Department, Geneva University Hospitals (HUG), Geneva, Switzerland.
[2] Department of Radiology and Medical Informatics, Faculty of Medicine, University of Geneva, Geneva, Switzerland.
[3] Pharmacy, Geneva University Hospital (HUG), and Institute of Pharmaceutical Sciences of Western Switzerland, School of Pharmaceutical Sciences, University of Geneva, Geneva, Switzerland.
[4] Division of Endocrinology, Diabetes and Metabolism, Department of Internal Medicine, Geneva University Hospitals (HUG), Geneva, Switzerland.
[5] Quality Office, Diagnostic Department, Geneva University Hospitals (HUG), Geneva, Switzerland.
[6] Clinique La Colline, Hirslanden, Geneva, Switzerland

[*] Corresponding author: Jamil Zaghir, PhD, Division of Medical Information Sciences, Diagnostic Department, Geneva University Hospitals (HUG), Rue Gabrielle Perret-Gentil 4, Geneva, Switzerland. Email: jamil.zaghir@hug.ch. Phone number: +41 79 768 69 70.

[‡] Equal contribution

## ABSTRACT

**Objectives**: Large language models (LLMs) are increasingly used for clinical text summarization, yet structured methods to assess associated patient safety risks remain limited. Failure Mode, Effects, and Criticality Analysis (FMECA) provides a proactive framework for systematic risk identification but has not been adapted to LLM-generated clinical content. This study aimed to develop and validate a novel FMECA framework for the prospective assessment of patient safety risks in LLM-generated clinical summaries.

**Materials and Methods**: An interdisciplinary expert panel (n = 8) developed a taxonomy of failure modes through literature review and brainstorming. Standard FMECA dimensions (occurrence, severity, detectability) were adapted into 5-point ordinal scales. The framework was applied to 36 discharge summaries from four patients, generated by an open LLM (GPT-OSS 120B) using real-world clinical data from the Geneva University Hospitals. Reviewers independently annotated the summaries across two rounds. Inter-rater reliability was assessed at failure mode, severity and detectability score levels. Usability and content validity were evaluated using an adapted System Usability Scale and structured feedback.

**Results**: The final framework comprised 14 failure modes organized into categories. Inter-rater agreement improved between rounds, reaching moderate-to-substantial agreement for failure mode identification and good agreement for severity and detectability scoring. Usability was rated as good (mean SUS: 79.2/100), with high evaluator confidence.

**Discussion and Conclusion**: This study presents the first FMECA-based framework for systematic patient safety risk assessment of LLM-generated clinical summaries. The framework provides a structured and reproducible method for identifying clinically relevant risks caused by these summaries.

# INTRODUCTION

The digitization of patient records has fragmented clinical information, contributing to increased documentation burden and reduced efficiency in care delivery [1,2]. Large language models (LLMs), characterized by their ability to process high-dimensional data and generate coherent narrative text, have emerged as transformative tools for clinical summarization. When applied to Electronic Health Records (EHRs), these models hold the potential to streamline workflows and enhance clinical decision support by synthesizing disparate data points into actionable insights [3–7].

However, as research into LLM-enabled summarization advances, the associated patient safety risks remain critically underdefined. Current literature frequently lacks systematic failure analysis, and the use of inconsistent terminology, such as the loosely defined concept of "hallucination", hampers the development of standardized error taxonomies [8]. Consequently, the frequency, severity, and downstream clinical consequences of LLM-generated errors remain insufficiently understood [9,10]. Such errors pose a substantial threat: if propagated through the clinical record, they may create latent patient safety hazards that are difficult to intercept.

This clinical risk is compounded by a burgeoning regulatory gap. While the European Union Artificial Intelligence Act (EU AI Act) and the U.S. Food and Drug Administration (FDA) have established guidelines for Software as a Medical Device (SaMD), many LLM applications currently operate in a "gray zone" [11,12]. Under the guidance of the Medical Device Coordination Group (MDCG) 2019-11 rev.1 (June 2025) [13], an LLM fundamentally meets the regulatory definition of software as a "set of instructions that processes input data and creates output data". When such a model is specifically intended for a medical purpose, such as summarizing clinical data to influence a physician's diagnostic or treatment pathway, it transcends its status as a general-purpose tool to become Medical Device Software (MDSW), the European regulatory equivalent of Software as a Medical Device (SaMD). According to Rule 11 of the Medical Device Regulation (MDR EU 2017/745) [14], any software providing information used to take clinical decisions is classified as Class IIa or higher, necessitating a formal conformity assessment.

In this high-stakes regulatory landscape, Failure Modes, Effects, and Criticality Analysis (FMECA) serves as the gold standard for proactive risk management, traditionally used in some medical areas such as pharmaceutical sciences [15–17] and medical devices [18,19] to identify potential failures before they reach the patient. However, current research suggests that many practitioners struggle to bridge the gap between academic risk theory and the rigid requirements of international standards like ISO 14971 [20,21], often treating FMECA as a "compliance-first" exercise rather than a driver of safety innovation. However, while FMECA is commonly applied to deterministic hardware or static software, there is, to our knowledge, no prior work that leverages it to represent the inherently stochastic behavior and probabilistic risk profiles of generative AI in clinical settings. Therefore, there is a clear need to develop an adapted FMECA framework that offers a structured, forward-looking approach capable of addressing the non-linear and unpredictable failure modes characteristic of clinical LLM applications.

To address this gap, we propose the first structured risk analysis framework specifically designed to evaluate patient safety risks associated with LLM-generated clinical summaries. This work introduces three key contributions: (i) we design an initial FMECA framework tailored to LLM-based clinical summarization through a consensus process, defining both a taxonomy of failure modes and associated scoring scales for occurrence, severity, and detectability; (ii) we operationalize and evaluate this framework on LLM-generated summaries derived from real-world discharge documents, enabling an empirical assessment of its applicability in practice; and (iii) we iteratively refine the framework based on insights and feedback from this evaluation, and provide a combined qualitative and quantitative analysis of its reliability and clinical relevance, including measures of inter-annotator agreement. To the best of the authors' knowledge, this study constitutes the first prospective application of FMECA to generative AI outputs in clinical practice, offering a scalable and systematic model for the safe integration of LLMs into the healthcare ecosystem.

# MATERIALS AND METHODS

## Rationale: Traditional FMECA

Traditional FMECA was selected for its structured, proactive approach to failure identification and risk prioritization. The probabilistic variant was excluded: it requires established failure rates, failure mode ratios, and conditional effect probabilities, none of which are available for LLM-based systems. Traditional FMECA, by relying on severity classification and expert judgment rather than statistical inputs, is better suited to this context.

## FMECA: Scope and Framework Design

The analysis focused exclusively on risks arising during the content generation phase of clinical summarization (**Figure 1**). This restriction isolates risks intrinsic to LLM generation process, excluding non-generative risks like data extraction or power supply issues.

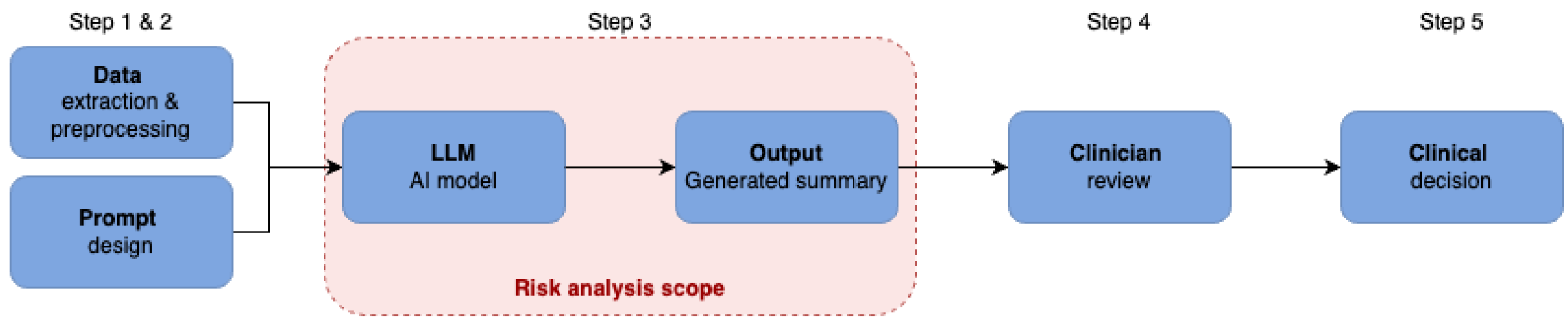

**Figure 1.** Process flow diagram of the LLM-assisted clinical summarization workflow. The workflow comprises five steps: data extraction and preprocessing (Step 1), prompt design (Step 2), data processing by the AI model (Step 3), clinician review (Step 4), and clinical decision-making (Step 5). The risk analysis focuses specifically on Step 3. Steps 1, 2, 4, and 5 are excluded from the analysis, as they involve processes external to the LLM's generative task.

An interdisciplinary panel was convened to design the FMECA framework. The panel included experts in FMECA & risk analysis (P.B., C.S.), clinical informatics (C.L.), large language models (J.Z. and L.B.), quality assurance (C.G.), and clinical practice (K.G., C.R-M.). The panel met twice: first to brainstorm potential failure modes, and second to refine definitions and adapt FMECA scoring scales.

Failure modes were identified through a three-step process. First, panelists independently listed potential failures using the prompt: "What could go wrong at this stage?" Additional coverage was ensured using previously reported failure modes from the literature [8]. Second, responses were consolidated and deduplicated. Third, a second meeting was conducted to refine, classify, and validate the failure modes. Items were excluded if they were out of scope, overly granular, or described root causes rather than observable failures.

Standard FMECA dimensions – occurrence, severity, and detectability – were designed to suit the specific context of the study. A 5-point ordinal scale was preferred over the often used 10-point scale to reduce inter-rater variability, consistent with established healthcare risk assessment practices [22].

## FMECA validation: Experimental Setup

### Data Collection

A retrospective study was conducted using clinical data from the EHRs of Geneva University Hospitals (HUG). All data processing and evaluation were performed within secure institutional infrastructure. The study was approved by the cantonal ethics committee (CCER Approval ID: 2025-01062).

Patient data for this study were derived from a larger ongoing cohort. In the parent study, all patients aged ≥18 years who provided informed consent for data use were included (Supplementary Material) Patients whose hospitalizations were exclusively related to uncomplicated pregnancies, as identified by ICD-10 codes depicted in Supplementary Material, were excluded. The final cohort comprised 846 patients: 465 females (55%; median age 55 years [IQR: 38–70]) and 381 males (45%; median age 70 years [IQR: 61–80]). The mean number of hospitalizations per patient was 10.36 (SD: 10.01).

The source cohort included adult patients with at least two hospitalizations after January 1, 2022. From this cohort, a purposive subsample of four patients (i.e. a total of 36 discharge summaries in PDF format) was selected to represent varying levels of clinical complexity, as approximated by the number of hospitalizations (2, 4, 12, and 19, respectively). Text was extracted from PDF files and preprocessed to remove non-informative elements (e.g., headers, signatures).

### LLM-based Summarization

The task consisted of generating a structured summary from individual hospital discharge summaries. Each document was independently provided to the LLM, which was prompted to extract and organize clinically relevant information into three predefined sections: (i) medical history, with active or resolved status and associated treatments; (ii) known allergies and reactions; and (iii) a summary of the current clinical episode, including the reason for consultation, diagnosis, management, and follow-up plan.

Due to data sensitivity, an open-weight model (GPT-OSS 120B [23]) was used and deployed locally. This model was chosen for its 128k-token context window, its feasibility on modest GPU hardware (requiring 70 GB of GPU RAM), and its state-of-the-art performance on established benchmarks, including MMLU Pro [24] and GPQA-Diamond [25]. All inferences were performed on an Apple M2 Ultra (76-core GPU, 24-core CPU, 192 GB unified memory), with the temperature set to 0.1 to favor precise and concise outputs.

Prompt design [26] combined persona prompting (clinician role) and template prompting (fixed output structure). The model was instructed to produce factual, concise summaries strictly grounded in the source text, with missing information explicitly noted. The full prompt is available in Supplementary Material.

### Evaluation Process using the Framework

Each generated summary was independently evaluated by at least two trained clinical reviewers. A total of three reviewers participated, all of whom were physicians: two with clinical experience in internal medicine, and one recently graduated physician. Reviewers compared outputs to source documents to identify failure modes from taxonomy. Multiple instances of the same failure could be recorded within a single summary.

For each identified failure, reviewers assigned two quantitative scores: severity, reflecting the potential clinical harm if the error went undetected, and detectability, reflecting how likely a clinician would be to notice the error during routine use of the summary.

Failure mode identification and scoring were performed using a purpose-built web-based annotation interface. The interface comprised three components: (1) a binary annotation grid, in which reviewers indicated the presence or absence of each failure mode for a given summary; (2) an open-text comment field, in which reviewers could provide a comment for each identified failure; and (3) a risk scoring module, in which reviewers assigned severity and detectability ratings according to the predefined scales.

The evaluation was conducted in two rounds:

- Round 1: Two reviewers annotated all summaries with minimal training. This exploratory phase identified ambiguities and usability issues.
- Round 2: After FMECA framework refinement based on Round 1, a third clinical reviewer was included. All three reviewers completed standardized training. The training comprised a detailed review of the taxonomy, guided annotations of representative summaries, and structured group discussion of borderline cases to achieve consensus in interpretation. Then, all three reviewers independently re-evaluated the summaries using the revised framework.

### Framework Validation: Quantitative metrics

The validation of the FMECA framework was performed in both rounds: the inter-annotator agreement was computed to evaluate the consistency of the framework across evaluators and to provide indirect evidence of the clarity and robustness of the taxonomy.

Stage 1 assessed binary subcategory-level agreement, and Stage 2 assessed binary failure mode-level agreement (presence/absence of each specific failure mode), where each subcategory comprises one or more failure modes. Finally, Stage 3 assessed agreement in severity and detectability ratings. For binary agreement (Stages 1-2), pairwise agreement used Cohen's kappa and Gwet's AC1; multi-rater agreement used Fleiss' kappa, Gwet's AC1, and Krippendorff's alpha. For ordinal ratings (Stage 3), pairwise agreement used Pearson's r and Spearman's ρ; multi-rater agreement used Intraclass Correlation – i.e. ICC(2,1). Agreement within ±1 and ±2 points on the 5-point scales was reported as a measure of practical concordance.

### Framework Validation: Usability and Qualitative Validation

Usability validation was performed using an adapted System Usability Scale (SUS) [27] (detailed SUS is available in Supplementary Material). In addition, structured qualitative feedback was collected to assess four dimensions of the framework: (1) cognitive load associated with the evaluation task, (2) interpretability of the taxonomy, including clarity and ambiguity of failure modes definitions and potential category overlap, (3) comprehensiveness of the taxonomy, including the identification of missing failure modes, and (4) the usability of the severity and detectability scales.

# RESULTS

## Risk-based Framework Development

### Failure Modes Identification

The initial phase combining literature review and brainstorming identified 78 candidate failure modes, which were reduced to 51 unique items after duplicate removal. The interdisciplinary experts further excluded 27 items (11 overly granular, 16 representing root causes), resulting in a preliminary taxonomy of 20 failure modes (available in the Supplementary Material).

Between Round 1 and Round 2, iterative refinement based on reviewer feedback and identified ambiguities led to consolidation of overlapping concepts, yielding an intermediate taxonomy of 15 failure modes. The Round 2 inter-rater agreement results reported below are therefore based on this 15-item taxonomy. Following completion of Round 2, two failure modes were identified as conceptually dependent and were merged, yielding the final taxonomy of 14 failure modes. Merging was driven by conceptual similarity (e.g., lexical, syntactic, and logical inconsistencies grouped under ambiguous formulation). No additional failure modes were identified, supporting the taxonomy's completeness. This led to a framework comprising top-level categories, subcategories, and failure modes (**Figure 2**).

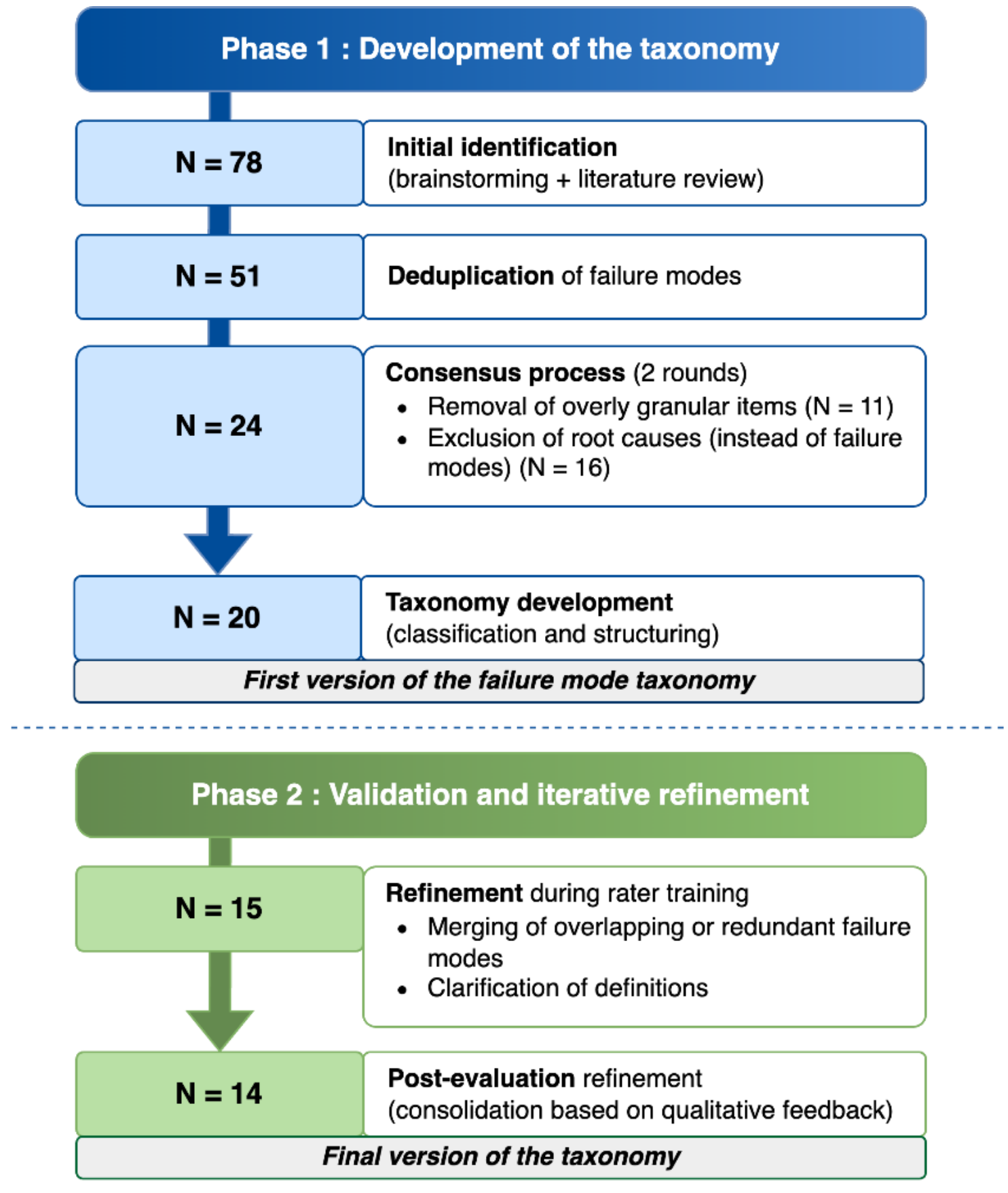

**Figure 2.** Flow diagram of failure mode identification and taxonomy development. Failure modes were identified (N = 78), deduplicated (N = 51), and refined through a consensus process (N = 24). Classification resulted in a preliminary taxonomy (N = 20), which was further reduced through iterative refinement during rater training and evaluation, yielding a final set of 14 failure modes.

As shown in **Table 1**, the final taxonomy is structured into six categories and ten subcategories, balancing granularity with usability.

## Scoring Scales Adaptation

Severity, defined as the potential of harm of an undetected failure mode, was adapted from the Agency for Healthcare Research and Quality (AHRQ) Common Format Harm Scale [28]. Scores range from 1 (no plausible clinical impact) to 5 (direct or indirect contribution to patient death) (**Table 2**).

Detectability, defined as the likelihood that a clinician would recognize a failure mode during routine use of the LLM-generated summary, was designed de novo for this study. The scale captures a gradient of detection difficulty based on two key dimensions: (1) whether the error can be identified from the summary alone or requires consultation of the source document, and (2) the cognitive effort and time needed for recognition. Scores range from 1 (immediately and universally obvious upon reading) to 5 (very unlikely to be detected before influencing clinical reasoning, even when the source document is available) (**Table 3**).

Occurrence was defined as the proportion of summaries in which a failure mode appeared, providing a stable, output-based measure of frequency. As depicted in **Table 4**, scoring thresholds were adapted from ratio-based method proposed by Asgari et al. [29].

**Table 1.** Hierarchical taxonomy of failure modes for LLM-generated clinical summaries. The taxonomy comprises five categories, nine subcategories, and specific failure modes. Illustrative manifestations of consolidated failure modes are shown in italics.

| Category | Subcategory | Failure Mode |
|---|---|---|
| Faithfulness to the Query | | |
| | Structure | • General structural error |
| | | • Information placed in an inappropriate section |
| | Content | • Addition of out-of-context information |
| | | • Inadequate level of detail *(e.g., too detailed or too vague)* |
| | Vocabulary | • Vocabulary inappropriate to the context |
| Readability | | |
| | Intelligibility | • Ambiguous formulation *(e.g., contradictions, logical breaks, lexical or syntactic errors affecting comprehension)* |
| | | • Response in the wrong language |
| | Conciseness | • Lexical redundancy |
| | | • Redundancy of medical information |
| Ethical Appropriateness | | |
| | | • Stigmatizing or discriminatory vocabulary |
| Faithfulness of Content Relative to the Source Document | | |
| | Factual fidelity to source document information | • Presence of factually incorrect information relative to the source document(s) *(e.g., date errors, errors in managing interrelated information)* |
| | Content traceability relative to the source document | • Presence of information absent from the source document(s) *(e.g., fabricated information, subjective interpretation)* |
| Exhaustivity | | |
| | | • Omission of information present in the source document(s) |
| Technical Issue | | |
| | Summary generation | • Failure to generate the summary |

**Table 2**. Severity rating scale.

| Score | Description | Definition |
|---|---|---|
| 1 | None | The failure mode has no plausible clinical impact on the patient or the care process, even if used in practice. |
| 2 | Minor | The failure mode could affect the patient but would not cause physical or psychological harm and would not require any medical intervention. |
| 3 | Considerable | The failure mode could cause reversible physical or psychological harm, requiring additional care or treatment, without major medical intervention. |
| 4 | Major | The failure mode could cause irreversible harm (permanent injury) or reversible harm requiring a major medical intervention (e.g., surgery, transfer to intensive care), without being immediately life-threatening. |
| 5 | Catastrophic | The failure mode could directly or indirectly contribute to the patient's death, whether immediate or delayed. |

Table 3. Detectability rating scale.

| Score | Description | Definition |
|---|---|---|
| 1 | Very easily detectable | The error is immediately and universally obvious upon reading the summary (<10 seconds), without requiring clinical expertise. |
| 2 | Easily detectable | The error is detectable from the summary alone after brief attention or reflection (≤1 minute), without consulting the source document or performing in-depth analysis. |
| 3 | Detectable but not immediate | The error is detectable from the summary alone, but only after careful reading, contextual reasoning, or prolonged examination (>1 minute); detection is not systematic and does not require consulting the source document. |
| 4 | Poorly detectable | The error is unlikely to be detected from the summary alone and can only be identified through a systematic review of the source document(s). |
| 5 | Very poorly detectable | The error is very unlikely to be detected before influencing clinical reasoning or patient care, even if the source document is available. |

Table 4. Occurrence rating scale.

| Score | Description | Definition |
|---|---|---|
| 1 | Very low | < 1% |
| 2 | Low | 1-10 % |
| 3 | Medium | 10 - 60 % |
| 4 | High | 60 - 90 % |
| 5 | Very high | > 90 % |

## Framework Validation

### Inter-annotator agreement

#### *Round 1*

In the initial two-rater round, subcategory-level agreement showed a marked discrepancy between Cohen's kappa (0.11, slight) and Gwet's AC1 (0.66, substantial), consistent with the well-known kappa paradox: low prevalence of flagged subcategories led to an artificially deflated kappa despite high observed agreement with AC1. A similar pattern emerged at the failure mode level (Stage 2), where Cohen's kappa was 0.15 (slight) and Gwet's AC1 reached 0.87 (almost perfect). Given the strong class imbalance, where non-flagging overwhelmingly predominates, improving chance-corrected agreement (Cohen's kappa) is more appropriate than relying on raw agreement (Gwet's AC1), which is largely driven by concordant non-flagged cases.

At the score level (Stage 3), pairwise correlations were moderate to high, with Pearson's $r = 0.493$ and Spearman's $\rho = 0.453$ for severity, and Pearson's $r = 0.646$ and Spearman's $\rho = 0.593$ for detectability. Agreement within ±1 point reached 80% for both dimensions, and 100% within ±2 points.

Following this round, residual disagreements were examined. Failure mode definitions were clarified and scoring anchors refined to reduce ambiguity.

#### *Round 2*

Round 2 was conducted using the refined 15-item taxonomy. As shown in **Figure 3**, subcategory-level agreement across three raters was moderate (Fleiss' kappa = 0.424, Gwet's AC1 = 0.515, Krippendorff's alpha = 0.425), with unanimous agreement in 60% of items. Pairwise analyses showed consistent results across rater pairs (Cohen's kappa: 0.35–0.48; Gwet's AC1: 0.50–0.54). At the failure mode level (Stage 2), multi-rater agreement was moderate to substantial (Fleiss' kappa = 0.400; Gwet's AC1 = 0.654), with 67% unanimous agreement. Notably, Cohen's kappa increased compared to the first round, improving from Slight to Moderate agreement and effectively bypassing the fair agreement range.

Score-level agreement (Stage 3) yielded strong agreements: the ICC(2,1) was 0.719 (good correlation) for severity and 0.731 (good correlation) for detectability. Exact agreement occurred in 39% of severity ratings and 47% of detectability ratings. Agreement within ±1 point reached 92% for severity and 84% for detectability, and within ±2 points reached 100% and 87%, respectively. These findings indicate that the adapted scoring scales were interpreted consistently across the three evaluators, supporting their reliability within the FMECA framework.

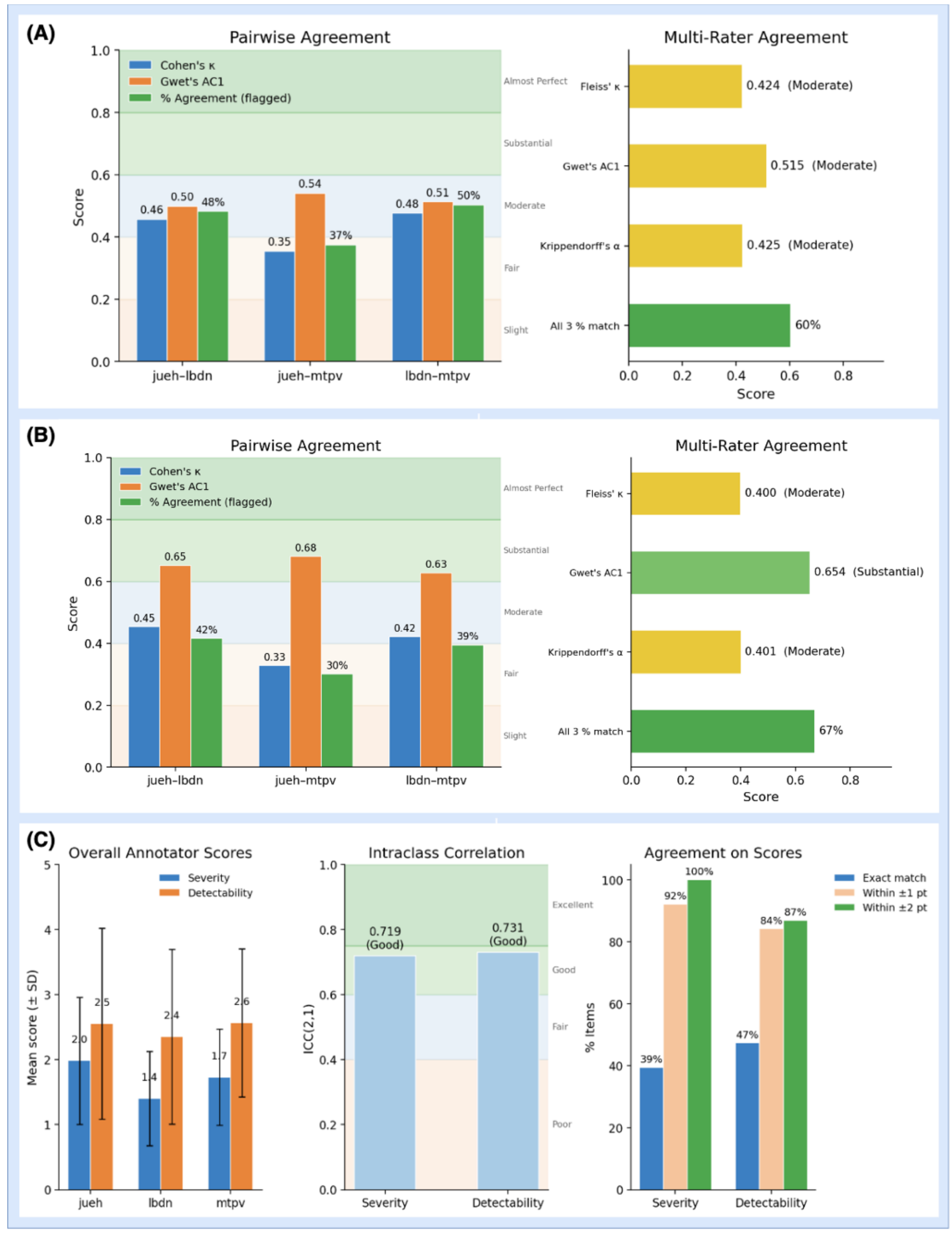

**Figure 3.** Inter-rater reliability of the adapted FMECA framework (Round 2, three raters). (A) Stage 1 - Agreement at the subcategory level. (B) Stage 2 - Agreement at the failure mode level. (C) Stage 3 - Agreement at the scoring scales level. Left: mean severity and detectability scores (± SD) by evaluator. Center: intraclass correlation coefficients (ICC(2,1), two-way random, single measures) for severity and detectability. Right: proportion of score pairs in exact agreement, within ±1 point, and within ±2 points on the 5-point scales.

## Qualitative Evaluation: Usability and Content Validity

Three evaluators completed the adapted SUS (Supplementary Material) after going through the 2-round framework validation, yielding a mean score of 79.2/100 (SD ± 6.3; individual scores: 87.5, 77.5, 72.5). This corresponds to a "Good" rating (Grade B+), just below the "Excellent" threshold of 80.3. All evaluators consistently indicated that the framework did not require technical support (Q4 ≤ 2), was not awkward to use (Q8 ≤ 2), and that its components were well integrated (Q5 ≥ 4). They also reported high confidence in their use of the framework (Q9 ≥ 4). The greatest variability was observed for Q10 (need for prior learning), with scores ranging from 2 to 4, reflecting differences in onboarding experience.

Structured feedback further supported these findings. All evaluators rated the severity and detectability scales as easy to interpret, citing their clearly defined intervals. One evaluator noted improved clarity following the initial refinement phase. Cognitive load was rated as moderate by two evaluators and high by one, the latter attributing this to the sustained attention and critical judgment required by the task. Two evaluators independently identified overlap between the failure modes “Presence of subjectivity or interpretation” and “Information fabrication”. No missing failure modes were reported. One evaluator further noted that ambiguity during annotation was primarily driven by the quality of the source data rather than limitations of the framework itself.

# DISCUSSION

## Summary of the Findings

This study presents, to our knowledge, the first use of a FMECA framework for the assessment of patient safety risks in LLM-generated clinical summaries. By combining a hierarchical taxonomy of failure modes with scoring scales, the framework enables systematic risk characterization. Across two validation rounds, the framework demonstrated satisfactory usability, strong score-level reliability, and moderate-to-substantial agreement for failure mode identification.

## Reliability of the framework

A central contribution of this study lies in its evaluation of inter-rater agreement. While initial agreement appeared slight (0.15) when assessed using Cohen’s kappa, it has improved following an iterative refinement (0.33-0.45, depending on the reviewer pair). These results must be interpreted considering the annotation task: for each summary, reviewers performed 15 independent binary decisions, one per failure mode, determining presence or absence without mutual exclusivity, meaning any combination could co-occur. Each decision required detailed comparative judgment between the generated summary and the source document, without a gold-standard reference, in linguistically fluent text that does not signal its own errors. When a failure mode was identified, reviewers additionally assigned severity and detectability scores on 5-point ordinal scales.

This combined detection, classification, and scoring task is considerably more demanding than those reported in comparable studies. Altermatt et al. [30] reported a mean Cohen’s kappa of 0.73 for error classification of GPT-4o outputs using a 7-category taxonomy. However, three features of their design inflate agreement relative to ours: outputs were guaranteed to contain an error, eliminating the detection step; reviewers were required to select exactly one category per item, reducing the task to forced-choice classification; and the taxonomy was half the size, mechanically lowering the probability of disagreement. Asgari et al. [29] whose framework most closely resembles ours, did not report formal inter-rater metrics at all.

Despite these challenges, several findings support the robustness of the framework. The introduction of a third independent reviewer in Round 2 ensures that the observed improvement reflects genuine gains in framework clarity rather than mere rater calibration. Notably, agreement was consistently higher for severity and detectability scoring (ICC > 0.7), indicating that once a failure mode is recognized, its clinical significance can be reliably assessed. Residual disagreement at the identification level may reflect both the inherent difficulty of detecting subtle errors in fluent clinical text and residual ambiguity in certain taxonomy definitions, which future iterations should address. Together, these findings suggest that, although perfect agreement remains unlikely given the interpretative nature of clinical risk identification, structured methodologies can mitigate variability.

## Methodological Considerations and Limitations of FMECA

While FMECA provides a proactive framework for risk assessment, its methodological limitations must be acknowledged. A key component of traditional FMECA is the Risk Priority Number (RPN), defined as the product of occurrence, severity, and detectability. In this study, however, the estimation of occurrence is constrained by the limited dataset (four patients and 36 summaries), precluding robust frequency estimation. Consequently, the occurrence and RPN scores require further validation through large-scale studies.

More broadly, the RPN itself has been widely criticized in the literature [31]. A systematic review of 236 FMECA studies [32] identified inherent deficiencies in the RPN formulation, including its mathematical properties and sensitivity to subjective scoring. The multiplicative structure assumes equal weighting and independence of dimensions, which may not reflect clinical reality, and can yield identical RPN values for qualitatively different risk profiles. These limitations are particularly relevant in the context of LLM evaluation, where scoring inherently involves subjective judgment.

Nevertheless, FMECA remains a widely used tool for structured risk identification, particularly in exploratory settings. In this study, the primary objective was not to derive definitive risk rankings, but to establish a reliable framework for identifying and characterizing failure modes. Future work may explore aggregation strategies to address limitations of the RPN.

## Positioning Within the Existing Literature

Several previous studies aimed to evaluate LLM performance in healthcare, including standardized benchmarks [33], and human expert assessment instruments such as the PDSQI-9 [34]. However, structured methods for the prospective identification of patient safety risks at the individual output level remain scarce [8,10].

Asgari et al. [29] proposed a framework for error detection in LLM-generated clinical notes using a binary taxonomy (hallucination vs. omission), with severity classified as major or minor and a risk matrix combining occurrence and severity. The present framework differs in two aspects.

First, it replaces coarse binary categories with a hierarchically structured taxonomy of 14 failure modes developed through a consensus process. Beyond enabling finer-grained characterization, this addresses a more fundamental limitation: terms such as "hallucination" lack a standardized definition in medical contexts [35], and ambiguous categories risk obscuring clinically distinct error types.

Second, the framework explicitly incorporates detectability as a standalone risk dimension, an addition grounded in FMECA methodology and particularly relevant for LLM-generated content, where errors may be linguistically fluent yet clinically misleading. Capturing detectability alongside severity yields a more complete representation of risk than severity-occurrence approaches alone.

## Study strengths and limitations

This study's framework was developed through a rigorous, interdisciplinary process and refined iteratively based on empirical findings. The multi-level reliability assessment provides a comprehensive evaluation of its measurement properties, and the inclusion of detectability represents a conceptual advance aligned with the unique characteristics of LLM-generated text.

Limitations include a sample size that constrains the generalizability of occurrence estimates and RPN-based prioritization. The evaluation was restricted to a single LLM, task, and language (French), and its transferability to other settings remains to be established.

Risk identification in healthcare is inherently incomplete: not all hazards can be anticipated, and novel failure modes may emerge as technologies evolve. As such, the proposed taxonomy should be viewed as dynamic and subject to ongoing refinement.

## Implications and Future Directions

This work demonstrates that FMECA can be adapted to the evaluation of LLM-generated clinical content, providing a structured approach to identifying patient safety risks. The findings show the importance of iterative calibration, standardized training, and clear operational definitions in achieving acceptable reproducibility.

Future research will focus on scaling the framework to larger datasets, enabling estimation of occurrence and more reliable risk prioritization. Additional work is also needed to validate the framework across different clinical tasks, models, and languages, and to explore alternative approaches to risk aggregation that address known limitations of the RPN.

# CONCLUSION

This study introduces a novel FMECA framework for the systematic identification and quantification of patient safety risks associated with LLM-generated clinical summaries. By combining an expert-derived taxonomy of 14 failure modes with tailored occurrence, severity, and detectability scales, the framework enables structured and transparent risk characterization. Reliability analyses demonstrated good inter-rater agreement, supporting the consistency and practical applicability of the approach, particularly after iterative refinement.

Although based on a limited sample and a single model, this work provides an important step toward standardized, prospective safety evaluation of generative AI in clinical settings. In contrast to performance-focused assessments, the proposed framework emphasizes clinically meaningful risk identification and prioritization, addressing a critical gap in current evaluation practices. As LLM-based tools move closer to real-world deployment, such structured methodologies will be essential to ensure that their integration into care processes maintains, rather than compromises, patient safety.

Future work should validate the framework across larger and more diverse datasets, multiple LLM architectures, and varied clinical applications. Extending the analysis to downstream clinical use and exploring hybrid human-automated evaluation strategies may further enhance scalability and impact.

## Acknowledgments

Thank you to interdisciplinary expert panel. The authors used AI for grammatical correction and rephrasing to improve clarity. The authors reviewed, edited the final content, and take full responsibility for the paper's accuracy.

## Conflicts of Interest

The authors declare that they have no known competing financial interests or personal relationships that could have appeared to influence the work reported in this paper.

## Funding

This work was funded with a grant from the Private Foundation of the Geneva University Hospitals.

## Author contributions

L.B. and J.Z. contributed equally to this work and were responsible for Conceptualization, Methodology, Data Curation & Extraction, Formal Analysis & Statistics, and Writing – Original Draft. L.B., J.E., and M.T. conducted the Framework Review & Validation. The Consensus Process involved L.B., J.Z., C.L., C.S., K.G., C.G., C.B.R., and P.B. Then, L.G. and R.D. contributed to Data Curation & Extraction and Formal Analysis & Statistics. M.B., C.G.B. and C.L. were responsible for Supervision and Project Management. All authors contributed to Writing – Review & Editing and approved the final version of the manuscript.

## Data Availability Statement

Patient data are not publicly available due to privacy and ethical restrictions, despite the ethics committee approval (CCER ID: 2025-01062). The prompt used in this study is provided in Supplementary Material to support transparency and reproducibility. Additional materials (e.g., evaluation templates or code) are available from the corresponding author upon request.

# Supplementary materials

## Déclaration de consentement pour l'utilisation des données de santé et des échantillons à des fins de recherche

J'accepte que mes données de santé et mes échantillons biologiques collectés durant les soins (consultations ambulatoires et hospitalisations) soient utilisés à des fins de recherche

Oui ☐ Non ☐

J'ai compris :

- les explications sur la réutilisation de mes données de santé et échantillons biologiques à des fins de recherche, détaillées dans l'information ci-dessus.
- que mes données personnelles sont protégées.
- que mes données et échantillons biologiques peuvent être utilisés dans des projets de recherche nationaux et internationaux, dans les secteurs public et privé.
- que les projets peuvent inclure des analyses génétiques sur mes échantillons, à des fins de recherche.
- que je peux être recontacté-e si des résultats pertinents me concernant sont mis en évidence.
- que ma décision est libre et n'a pas d'effet sur mon traitement médical.
- que ma décision est valable pour une durée illimitée.
- que je peux retirer mon consentement à tout moment sans avoir à justifier ma décision.

Nom, prénom, date de naissance
ou étiquette patient

Lieu, date — Signature du patient, si capable de discernement

Lieu, date — Signature du représentant légal, si nécessaire (nom et relation avec le/la patient-e)

Si vous avez des questions, ou si vous souhaitez recevoir une copie de ce formulaire avec signature, prenez contact avec le **Centre de Recherche Clinique**
rue Gabrielle-Perret-Gentil 4, 1211 Genève 14
022 372 91 57, aider-la-recherche@hcuge.ch

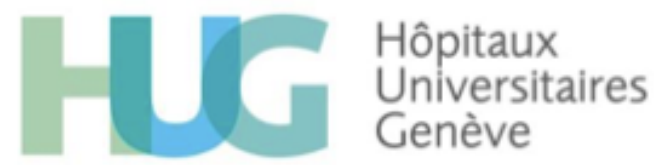

Septembre 2019

Supplementary Figure S1. Geneva University Hospitals consent form for research use of health data and biological samples.

# System Usability Scale (SUS)

Adapted for the FMECA Evaluation Framework
for LLM-Generated Clinical Summary Assessment

*Brooke, J. (1996). SUS: A 'Quick and Dirty' Usability Scale. Usability Evaluation in Industry, 189-194.*

**Evaluator Information**

Evaluator ID: ____________________ Date: ______ / ______ / __________

Role / Specialty: ______________________________________________

**Instructions**

Please rate each statement below based on your experience using the FMECA evaluation framework (taxonomy, scoring scales, and annotation interface) for assessing LLM-generated clinical summaries.

For each statement, circle the number that best reflects your level of agreement.
Please respond to all items, even if you feel an item is not entirely applicable.
In such cases, select the center point (3).

| |  | Strongly disagree 1 | 2 | 3 | 4 | Strongly agree 5 |
|---|---|---|---|---|---|---|
| 1. | I think that I would like to use this evaluation framework frequently. | ○ | ○ | ○ | ○ | ○ |
| 2. | I found the evaluation framework unnecessarily complex. (-) | ○ | ○ | ○ | ○ | ○ |
| 3. | I thought the evaluation framework was easy to use. | ○ | ○ | ○ | ○ | ○ |
| 4. | I think that I would need the support of a technical person to be able to use this evaluation framework. (-) | ○ | ○ | ○ | ○ | ○ |
| 5. | I found the various components of this evaluation framework (taxonomy, scales, interface) were well integrated. | ○ | ○ | ○ | ○ | ○ |
| 6. | I thought there was too much inconsistency in this evaluation framework. (-) | ○ | ○ | ○ | ○ | ○ |
| 7. | I would imagine that most clinicians would learn to use this evaluation framework very quickly. | ○ | ○ | ○ | ○ | ○ |
| 8. | I found the evaluation framework very awkward to use. (-) | ○ | ○ | ○ | ○ | ○ |
| 9. | I felt very confident using the evaluation framework. | ○ | ○ | ○ | ○ | ○ |
| 10. | I needed to learn a lot of things before I could get going with this evaluation framework. (-) | ○ | ○ | ○ | ○ | ○ |

Adapted from Brooke, J. (1996). SUS: A 'Quick and Dirty' Usability Scale.
Geneva University Hospitals (HUG) — Division of Medical Information Sciences

Supplementary Figure S2. First page of the System Usability Scale (SUS) Adapted for the FMECA Evaluation Framework for LLM-Generated Clinical Summary Assessment.

### Scoring (for researcher use only)

Odd items (1, 3, 5, 7, 9): score contribution = response − 1
Even items (2, 4, 6, 8, 10): score contribution = 5 − response

SUS score = (sum of all 10 contributions) × 2.5     Range: 0–100

Interpretation: > 80.3 = Excellent (Grade A) | 68 = Average (Grade C) | < 51 = Poor (Grade F)

**SUS Score: __________ / 100**

### Additional Comments (optional)

Please provide any additional feedback on the usability of the evaluation framework:

### Structured Feedback

1. Were any failure mode definitions unclear or ambiguous? If so, which ones?

2. Were any failure modes missing from the taxonomy? If so, please describe.

3. Were any failure modes redundant or overlapping? If so, which ones?

4. Were the severity and detectability scales easy to interpret and apply?

5. How would you rate the cognitive load of the evaluation task? (low / moderate / high)

6. Any other suggestions for improving the framework or the annotation process?

Adapted from Brooke, J. (1996). SUS: A 'Quick and Dirty' Usability Scale.
Geneva University Hospitals (HUG) — Division of Medical Information Sciences

Suppelementary Figure S3. Second page of the SUS report form.

Supplementary Table S1. First version of the failure mode taxonomy, developed after the consensus process (N=20).

| Category | Subcategory | Failure Mode |
|---|---|---|
| Faithfulness to the Query | | |
| | Structure | • General structural error<br>• Information placed in an inappropriate section |
| | Content | • Addition of out-of-context information<br>• Inadequate level of detail *(e.g., too detailed or too vague)*<br>• Presence of subjectivity or interpretation |
| | Vocabulary | • Vocabulary inappropriate to the context |
| Readability | | |
| | Intelligibility | • Ambiguous formulation<br>• Inconsistencies in generated content (contradictions, logical breaks)<br>• Response in the wrong language<br>• Lexical errors affecting comprehension<br>• Syntactic errors affecting comprehension |
| | Conciseness | • Lexical redundancy<br>• Redundancy of medical information |
| Ethical Appropriateness | | |
| | | • Stigmatizing or discriminatory vocabulary |
| Faithfulness of Content Relative to the Source Document | | |
| | Factual fidelity to source document information | • Presence of factually incorrect information relative to the source document(s)<br>• Date errors<br>• Errors in managing interrelated information |
| | Content traceability relative to the source document | • Presence of information absent from the source document(s) |
| Exhaustivity | | |
| | | • Omission of information present in the source document(s) |
| Technical Issue | | |
| | Summary generation | • Failure to generate the summary |

Supplementary Table S2. List of ICD-10 codes used to exclude patients with exclusively uncomplicated pregnancies.

| ICD-10 Code | Description |
|---|---|
| O04.9 | Medical abortion, complete or unspecified, without complication |
| O47.1 | False labor at or after 37 completed weeks of gestation |
| O70.0 | First-degree perineal laceration during delivery |
| O70.1 | Second-degree perineal laceration during delivery |
| O75.7 | Vaginal delivery following previous cesarean section |
| O80 | Single spontaneous delivery |
| O80.0 | Single spontaneous delivery |
| O82 | Single delivery by cesarean section |
| O82.0 | Elective cesarean delivery |
| O92.5 | Suppression of lactation |
| O92.50 | Suppression of lactation |

Tu es un assistant médical chargé de résumer une lettre de sortie hospitalière selon un format structuré et clinique.

Objectif : Produire un résumé clair, factuel et concis des antécédents médicaux, allergies et de l'épisode clinique actuel, en respectant le modèle ci-dessous.

Une lettre de sortie à analyser :
**[One discharge summary expected here]**

Instructions de sortie:

1. Analyse soigneusement le texte.
   - Identifie les antécédents médicaux pertinents (diagnostics actifs ou résolus).
   - Repère les allergies et réactions éventuelles.
   - Résume l'épisode clinique actuel (motif, diagnostic, plan).

2. Présente le résultat dans le format EXACT suivant :

Antécédents médicaux

- [Antécédents 1] : [statut : actif / résolu / incertain] — [traitement ou commentaire pertinent]
- [Antécédents 2] : [statut : actif / résolu / incertain] — [traitement ou commentaire pertinent]
- …

Présente d'abord les antécédents médicaux actifs, puis résolus. S'il n'y en a pas, indiquer « Non mentionné ».

Allergies

- [Allergène 1] : [réaction ou effet]
- [Allergène 2] : [réaction ou effet]
- …

S'il n'y en a pas, indiquer « Non mentionné »

Résumé de la consultation

- Mme/M. [Nom si mentionné] a été consulté(e)/hospitalisé(e) pour [raison principale].
- Diagnostic(s): [diagnostic(s) retenu].
- Prise en charge : [traitements reçus, interventions].
- Plan de suivi : [suivi prévu].

Si des informations manquent, indiquer « Non mentionné ».

Contraintes de style:

- Utilise des phrases brèves, claires et médicalement neutres.
- Ne fais pas de déduction ou d'interprétation non mentionnée.
- Garde un ton professionnel et objectif.
- N'ajoute aucun commentaire, ou remarque personnelle.
- Si aucune information n'est disponible pour une section, écrivez « Non mentionné ».
- Utilisez un langage neutre et clinique.

Sortie attendue : uniquement le texte structuré selon le modèle ci-dessus.

Supplementary Figure S4. Original prompt used for summary generation (in French).

You are a medical assistant tasked with summarizing a hospital discharge summary in a structured, clinical format.

Objective: Produce a clear, factual, and concise summary of the patient's medical history, allergies, and current clinical presentation, following the template below.

A discharge summary to analyze:
**[One discharge summary expected here]**

Discharge instructions:

1. Carefully analyze the text.
    - Identify relevant medical history (active or resolved diagnoses).
    - Note any allergies and reactions.
    - Summarize the current clinical episode (reason, diagnosis, plan).
2. Present the result in the EXACT format below:

Medical History

- [History 1]: [status: active / resolved / uncertain] — [relevant treatment or comment]
- [History 2]: [status: active / resolved / uncertain] — [relevant treatment or comment]
- …

List active medical histories first, followed by resolved ones. If there are none, indicate "Not mentioned".
Allergies

- [Allergen 1]: [reaction or effect]
- [Allergen 2]: [reaction or effect]
- …

If none, indicate "Not mentioned".

Consultation Summary

- Ms./Mr. [Name, if provided] was seen/admitted for [main reason].
- Diagnosis(es): [diagnosis(es)].
- Treatment: [treatments received, procedures performed].
- Follow-up plan: [planned follow-up].

If information is missing, indicate "Not mentioned".

Style guidelines:

- Use short, clear, and medically neutral sentences.
- Do not make any unstated inferences or interpretations.
- Maintain a professional and objective tone.
- Do not add any comments or personal remarks.
- If no information is available for a section, write "Not mentioned."
- Use neutral, clinical language.

Expected output: only text structured according to the template above.

Supplementary Figure S5. Translated version of the prompt for readers.